\documentstyle[12pt,epsfig]{article}

\newlength{\dinwidth}
\newlength{\dinmargin}
\def\lapproxeq{\lower .7ex\hbox{$\;\stackrel{\textstyle
<}{\sim}\;$}}
\def\gapproxeq{\lower .7ex\hbox{$\;\stackrel{\textstyle
>}{\sim}\;$}}

\def\be{\begin{equation}}
\def\ee{\end{equation}}
\def\bea{\begin{eqnarray}}
\def\eea{\end{eqnarray}}

\def\msb{\overline{\rm MS}}

\begin{document}
\titlepage
\begin{flushright}
IPPP/02/01 \\
DCPT/02/02 \\
CERN-TH/2001--362 \\
Cavendish-HEP-2001/15 \\
January 2002 \\
\end{flushright}

\vspace*{2cm}

\begin{center}
{\Large \bf NNLO global parton analysis} \\

\vspace*{1cm}
\renewcommand{\thefootnote}{\fnsymbol{footnote}}
A.D. Martin$^a$, R.G. Roberts$^b$, W.J. Stirling$^a$ and R.S.
Thorne$^c$\footnote[2]{Royal Society University Research Fellow} \\

\vspace*{0.5cm} $^a$ Department of Physics and Institute for
Particle Physics Phenomenology, University of Durham, Durham, DH1 3LE \\
$^b$ Theory Division, CERN, 1211 Geneva 23, Switzerland. \\
$^c$ Cavendish Laboratory, University of Cambridge, Madingley
Road, Cambridge, CB3 0HE \\

\end{center}

\vspace*{1cm}

\begin{abstract}
We perform a NNLO (and a LO) global parton analysis in which we include the new precise
data for deep inelastic scattering from HERA and for inclusive jet production at the
Tevatron, together with the improved knowledge of the three-loop splitting functions.
The results are compared with our recent NLO analyses. The LO fit produces
significantly worse results in general, but gives a surprisingly good fit to the
Tevatron high-$E_T$ jet data. For the approximate NNLO analysis we notice a slight
improvement in the quality of the global fit, and find that the partons are changed by
up to $10\%$ at $Q^2=10\; {\rm GeV}^2$, in particular in the $x<0.01$ regime.
\end{abstract}

\newpage
\renewcommand{\thefootnote}{\arabic{footnote}}

The global parton analyses of deep inelastic scattering (DIS) and related hard
scattering data are generally performed at NLO order. Recently much effort has been
invested in computing NNLO QCD corrections to a wide variety of partonic processes
\cite{NNLOjunk} and therefore we need to generate parton distributions also at NNLO so
that the theory can be applied in a consistent manner.  Analyzing DIS at NNLO is
important in itself in that we may be able to investigate the hierarchy LO
$\rightarrow$ NLO $\rightarrow$ NNLO in the process where there are perhaps the most
precise data. Recently, there has been a significant increase in the precision of the
structure function data from the HERA experiments \cite{H1A}--\cite{ZEUSc} and these,
together with the inclusive jet production measurements at the Tevatron \cite{D0,CDF},
were used to generate an updated version of the NLO partons
\cite{MRST2001}.\footnote{As in \cite{MRST2001}, we use the D0 data obtained using the
cone algorithm rather than those obtained \cite{D0kt} using the $k_T$ algorithm, since
the former have greater rapidity coverage and have been studied more extensively.}

To carry out the analogous analysis at NNLO we need both the relevant splitting
functions as well as the coefficient functions. However, although the deep inelastic
coefficient functions are known at NNLO \cite{CF}, the splitting functions are not yet
available at this order.  On the other hand, valuable partial information is known in
the form of the few lowest moments of the splitting functions \cite{SF}, which tightly
constrain the large $x$ behaviour.  When this is combined with the known most singular
$\log 1/x$ behaviour \cite{S47}, it greatly limits the possible behaviour of the
functions down to quite small values of $x$. Indeed, van Neerven and Vogt \cite{VV12}
constructed a range of compact analytic expressions for the splitting functions that
are all compatible with this information.  In particular, they  provided closed
expressions which represent the fastest and the slowest permissible evolution. In an
exploratory NNLO global analysis of the data \cite{NNLO}, we, in turn, used these
expressions to obtain bands representing the spread of allowed parton distributions.
For example, the range of uncertainty of the NNLO gluon distribution was found to be
about $\pm 30\%$ ($\pm 15\%$) at $x \sim 10^{-4} (10^{-3})$.

Subsequently, the moment calculation of the three-loop splitting
functions was extended to give two higher moments \cite{RV}.  This
has allowed van Neerven and Vogt \cite{VV3} to provide analytic
expressions for the NNLO splitting functions which considerably
increase the reliability of the evolution in the region $x
\gapproxeq 10^{-3}$.  In other words, the allowed
bands of permissible splitting functions has shrunk considerably, even,
somewhat surprisingly, at small $x$.\footnote{The allowed band contains the
vast majority of permissible behaviours, but it is still possible to find
perfectly acceptable forms which lie considerably outside this region,
particularly at very small $x$; i.e. the bands
should not be regarded as anything like one-sigma errors in a
Gaussian distribution.}  The splitting functions are located near
the `slow evolution' edge of their previous bands.

Here, we use these improved constraints on the splitting functions to perform a new
NNLO global parton analysis of the available data sets, which include the new more
precise structure function measurements from HERA \cite{H1A}--\cite{ZEUSc} and the
inclusive jet data from the Tevatron \cite{D0,CDF}.  The analysis uses the same form of
the starting parameterizations and the same data sets \cite{H1A}--\cite{CDF},
\cite{CCFR}--\cite{BCDMSd}, with the same $Q^2$ and $W^2$ cuts, as our recent NLO
global analysis \cite{MRST2001}. In particular, we fit to DIS data with $Q^2 > 2\;{\rm
GeV}^2$ and $W^2 > 12.5\;{\rm GeV}^2$.\footnote{As for the NLO fit, we have
investigated the consequences of changing these cuts, and introducing cuts at small
$x$. The results are similar to NLO, and will be reported in \cite{cuts}.} In going to
a NNLO analysis, we use the three-loop expression for $\alpha_S$, in the $\overline{\rm
MS}$ scheme, and we extend the matching conditions when evolving through the heavy
quark thresholds. The details of the heavy quark prescription are very similar to those
in \cite{NNLO}, but we add a brief discussion of important points in the Appendix.
Although the deep inelastic data are described at NNLO, some other processes are not
yet calculated to this accuracy, and  so we have to use NLO expressions to fit to the
data for (i) Tevatron jet production, (ii) the charged lepton rapidity asymmetry from
$W$ boson hadroproduction and (iii) the Drell-Yan process\footnote{For Drell-Yan
scattering the inclusive rate is known at NNLO, but not the differential distribution
in $x_F$ that we use in the analysis.}. We note that the effect of the threshold
corrections at NNLO to Tevatron jet production has been calculated in \cite{owens}, and
this reduces scale uncertainty and perhaps is a good approximation to the full NNLO
correction. The result is approximately a $3-4\%$ increase in normalization of the
theory prediction (not surprising since the NLO correction is itself about a $7-10\%$
raise, largely independent of $E_T$), which would be completely subsumed in the
normalization error of the data.

The optimum global NNLO fit is obtained with the starting parameterizations of the
partons at $Q_0^2 = 1\;{\rm GeV}^2$ given by
\begin{eqnarray}
\label{eq:a2}
 x u_V & = & 0.262 x^{0.31} (1-x)^{3.50}(1+3.83x^{0.5}+37.65x) \\
\label{eq:a3}
 x d_V & = & 0.061 x^{0.35} (1-x)^{4.03}(1+49.05x^{0.5}+8.65x) \\
\label{eq:a4}
 x S & = & 0.759 x^{-0.12} (1-x)^{7.66}(1-1.34x^{0.5}+7.40x)\\
\label{eq:a5}
 x g & = & 0.669 x^{0.00} (1-x)^{3.96}(1+6.98x^{0.5}-3.63x)
-0.23 x^{-0.27} (1-x)^{8.7}.
\end{eqnarray}
The flavour structure of the light quark sea is taken to be
\begin{equation}\label{eq:a6}
  2\bar{u}, 2\bar{d}, 2\bar{s} \; = \; 0.4S - \Delta, \quad 0.4S +
  \Delta, \quad 0.2S
\end{equation}
with $s = \bar{s}$, as implied by the NuTeV data \cite{NuTeV}, and
where
\begin{equation}\label{eq:a7}
  x \Delta \; = \; x (\bar{d} - \bar{u}) \; = \;
1.432 x^{1.24} (1-x)^{9.66}(1+9.86x-29.04x^2).
\end{equation}
The masses of the quarks are taken to be $m_c =1.43\;{\rm GeV}$ and $m_b = 4.3\;{\rm
GeV}$, the former giving the best fit to the charm structure function data. The value
of $\alpha_S(M_Z^2)=0.1155$, and is displayed in Table~1. This fit corresponds to the
average of two extreme choices for the sets of splitting functions in Ref.~\cite{VV3}
and we can explore the uncertainty of the NNLO evolution by instead choosing either
extreme. These `slow' and `fast' evolutions we denote by A and B respectively.

For the sake of studying the progression from LO through NLO to NNLO, we also perform a
pure LO fit, for which the starting parameterization is
\begin{eqnarray}
\label{eq:lo1}
 x u_V & = & 0.474 x^{0.30} (1-x)^{3.12}(1-1.32x^{0.5}+19.56x) \\
\label{eq:lo2}
 x d_V & = & 0.668 x^{0.43} (1-x)^{4.03}(1-0.83x^{0.5}+7.68x) \\
\label{eq:lo3}
 x S & = & 0.458 x^{-0.19} (1-x)^{7.51}(1+0.025x^{0.5}+7.63x)\\
\label{eq:lo4}
 x g & = & 3.08 x^{0.10} (1-x)^{6.49}(1-2.96x^{0.5}+9.26x)
\end{eqnarray}
with
\begin{equation}\label{eq:lo5}
  x \Delta \; = \; x (\bar{d} - \bar{u}) \; = \;
4.163 x^{1.76} (1-x)^{9.51}(1+7.20x-24.8x^2).
\end{equation}
In this case  $\alpha_S(M_Z^2)=0.130$ (again see Table~1), and the parameterization of
the gluon is simpler because the fit no longer requires a negative input gluon at small
$x$ at this order.

\begin{table}[!h]
\caption{The QCD coupling, and corresponding $\Lambda$ parameter for $n_f = 4$ (in the
$\overline{\rm MS}$ scheme with $\mu^2=Q^2$ in all cases), for the LO, NLO and NNLO
fits. Note that as we progress from order to order a very different value of
$\Lambda_{\rm QCD}$ is needed to give the same $\alpha_S(M_Z^2)$, that is
$\alpha_S(Q^2)$ is defined differently at different orders.}
\begin{center}
\begin{tabular}{|l|c|c|} \hline
& $\alpha_S (M_Z^2)$ & $\Lambda_{\rm QCD}$ (MeV) \\ \hline
 LO & 0.130 & 220\\
 NLO & 0.119 & 323 \\
 NNLO & 0.1155 & 235\\ \hline
\end{tabular}
\end{center}
\end{table}

The quality of the NNLO fit to the major data sets is shown in Table~2, together with
that of the NLO \cite{MRST2001} and LO analyses.  For each of the smaller data sets,
e.g.\ the CDF $W-$asymmetry \cite{Wasymm} and E866 Drell-Yan asymmetry \cite{E866}, the
$\chi^2$ per degree of freedom is about 1 per point.  For all the DIS data sets the
numbers are quoted for statistical and systematic errors added in
quadrature.\footnote{A discussion of the influence of the correlated systematic errors
for the HERA data may be found in the Appendix of \cite{MRST2001}.}  The quality of the
fits to the individual data sets is satisfactory. For the E605 data the systematic
errors are quoted in a slightly ambiguous manner, and are generally subdominant, and so
we fit to statistical errors alone.  Hence the quite large $\chi^2$ in this case.  Our
treatment of the correlated systematic errors for the Tevatron jet data has been
discussed in Ref.~\cite{MRST2001}. In Fig.~1 we also show the LO, NLO and NNLO
descriptions of the $F_2$ data in a few representative $x$ bins.

\begin{table}[htb]
\caption{Quality of the NNLO fits to different data sets, together with the
corresponding $\chi^2$ values for the NLO and LO fits to the same data.}
\begin{center}
\begin{tabular}{|lccccc|} \hline
Data set & No.\ of & LO & NLO & NNLO & NNLOJ\\
& data pts & & & &\\ \hline
 H1 $ep$ & 400 & 493 & 382 & 385 & 381\\
 ZEUS $ep$ & 272 & 296 & 254 & 264 & 275\\
 BCDMS $\mu p$ & 179 & 179 & 193 & 173 & 175\\
 BCDMS $\mu d$ & 155 & 206 & 218 & 231 & 226\\
 NMC $\mu p$ & 126 & 131 & 134 & 121 & 120\\
 NMC $\mu d$ & 126 & 98 & 100 & 88 & 86\\
 SLAC $ep$ & 53 & 83 & 66 & 59 & 65\\
 SLAC $ed$ & 54 & 98 & 56 & 59 & 61\\
 E665 $\mu p$ & 53 & 50 & 51 & 55 & 59\\
 E665 $\mu d$ & 53 & 62 & 61 & 62 & 64\\
 CCFR $F_2^{\nu N}$ & 74 & 116 & 85 & 80 & 78\\
 CCFR $F_3^{\nu N}$ & 105 & 126 & 107 & 110 & 126\\
 NMC $n/p$ & 156 & 157 & 155 & 150 & 148 \\
 E605 DY & 136 & 282 & 232 & 220 & 268 \\
 Tevatron Jets & 113 & 123 & 170 & 186 & 124 \\ \hline
 Total & 2097 & 2500 & 2328 & 2323 & 2341 \\ \hline
\end{tabular}
\end{center}
\end{table}

For the fit to the DIS data the story is relatively straightforward. In general, as we
proceed from the LO$\rightarrow$NLO$\rightarrow$NNLO analysis, there is a sequential
improvement in the quality of the fit (with a couple of dissenting data sets), with the
incremental improvement being smaller at the second stage. The LO fit fails because the
evolution is a little too slow at small $x$ and also too slow, and not quite the
correct shape, at large $x$. This is because there are higher-order but
small-$x$-enhanced terms missing in the quark-gluon splitting function and higher-order
but large-$x$-enhanced terms missing in the coefficient functions. This can not be
completely countered even with a very large value of the coupling constant and a large
small $x$ gluon. The improvement in the NNLO fit is due to the inclusion of these small
and large $x$ terms at NNLO, and this also allows a smaller coupling constant than for
the NLO fit. However, the improvement in the quality of the fit to DIS data is not as
dramatic as that seen in Ref.~\cite{NNLO}. This is partially due to our raised $W^2$
cut -- NNLO working a little better in the now excluded $10\; {\rm GeV}^2 < W^2 <
12.5\; {\rm GeV}^2$ range. However, the improvement at small $x$ is not as great, and
this seems to be due to the fact that the fit actually preferred the previous estimate
of the NNLO splitting functions which had quicker small $x$ evolution. This may be a
sign that yet higher orders are important in this regime. For the Drell-Yan data the
pattern is the same as for DIS data, the improvement at each successive order being
correlated largely with the reduction in the coupling constant, since as we discovered
in \cite{MRST2001} the Drell-Yan data prefer a lower value of $\alpha_S(M_Z^2)$.

The situation for the Tevatron high-$E_T$ jet data is rather different and very
interesting. As we see from Table~2 there is actually a distinct deterioration in the
quality of fit to the jet data as we go to higher orders, and at LO we obtain almost as
good a fit as seems possible given the scatter of points. This is a very surprising
result (we are not aware it has been noticed previously), but actually quite easy to
understand. The way in which the different parton scattering combinations usually
contribute to the jet rate is shown in Fig.~2 of \cite{nigel}. At the high $E_T$ end of
the jet measurements the rate is mainly determined by quarks and is given roughly by
$\alpha_S^2(qq+q\bar q)$. In a standard fit this contribution is usually determined
very accurately and the normal excess in data must be accounted for by an increase in
the $g(q+\bar q)$ contribution, i.e. in the high $x$ gluon. Hence the better
description obtained in the special CTEQHJ \cite{CTEQ} and MRST2001J \cite{MRST2001}
sets. However, at high $x$ the quark coefficient functions are enhanced at large $x$ at
higher orders, behaving roughly like $\alpha_S\bigl(\ln(1-x)/(1-x)\bigr)_+$ at NLO, and
like $\alpha^2_S\bigl( (\ln(1-x))^3)/(1-x)\bigr)_+$ at NNLO. These positive
contributions lead to a smaller quark distribution at high $x$, as one goes to higher
order, in order to fit the DIS data. Also, as we have seen, the coupling required by
the best fit decreases with increasing order. The net effect is that
$\alpha_S^2(qq+q\bar q)$ at high $x$ is much (up to $40\%$) larger at LO than at NLO.
At $x\sim 0.1$, which is relevant for the lower $E_T$ end of the jet data, where the
rate is dominated by the $g(q+\bar q)$ and $gg$ contributions, 
the gluons are a little smaller at LO due
to the need to have more gluons at very small $x$, and this compensates for the
increased $\alpha_S$. Overall, this results in the shape of the parton distributions at
LO being very suitable for the description of the Tevatron jet data without any strong
constraints (the gluon at LO is naturally not too small at high $x$). The comparison of
the LO and NLO descriptions of the CDF jet data is shown in Fig.~2, and as one can see
the shape and normalization of the LO theory prediction compares very well to the data
without recourse to the large shifts in data from correlated systematic errors. As
shown in Table~2, this translates into a huge reduction in $\chi^2$. We find this
result of a very good match at LO, and a big change in going to NLO, very intriguing. This observation may possibly be relevant for analyses which 
extract the jet rates using LO
Monte Carlos with LO partons.\footnote{We have checked that the good jet fit is a
feature of LO partons in general, not just this particular set.}

Conversely at NNLO the quarks are slightly smaller at high $x$  than at NLO and this
means that there must be slightly more gluon needed at high $x$ for a good fit to the
jet data. Within the context of the global fit this means that at NNLO the fit to
Tevatron jet data is slightly worse and there is an even stronger constraint on the
high $x$ gluon than at NLO.\footnote{In \cite{NNLO} we did not incorporate independent
fits to the jet data at each order, i.e. the same high $x$ gluon constraint
was used for the NNLO as for the NLO analysis and that at LO was different
only by a constant K-factor. 
This is responsible in part for some of the differences
between the results in \cite{NNLO} and this paper, e.g. the higher value of
$\alpha_S(M_Z^2)$ used at LO and perhaps the smaller improvement in the quality of the
fit at NNLO compared to NLO.} This increase in the $\chi^2$ for the jet data at NNLO
means that the overall improvement in the quality of the global fit compared to NLO is
only marginal.

As in our NLO analysis \cite{MRST2001}, we have investigated the sensitivity to
$\alpha_S(M_Z^2)$. The results are similar, with an approximate error of $0.002$ again
being attributable due to the quality of the fit. The $\chi^2$ profile is similar to
that for the global fit in Fig.~16 of \cite{MRST2001}, except that the minimum is now
at $\alpha_S(M_Z^2)=0.1155$ and the increase in $\chi^2$ is slightly steeper for
increasing $\alpha_S(M_Z^2)$ rather than decreasing $\alpha_S(M_Z^2)$. Indeed the
profiles are rather similar to those in \cite{MRST2001} for each data set. As in
\cite{MRST2001}, where more details of the procedure can be found, we have also
obtained a special set of partons (denoted NNLOJ) by forcing a very good fit to the jet
data. The quality of this fit is shown in Table~2. As at NLO the coupling is required
to rise to obtain the best (NNLOJ) fit, but in this case only to
$\alpha_S(M_Z^2)=0.118$. The shape of the NNLOJ high-$x$ gluon is very similar to that
at NLO, seen in Fig.~15 of \cite{MRST2001}, and the way in which the global fit is
affected is similar to NLO, i.e. the main casualty of the improved jet fit is the
Drell-Yan comparison.

We now consider the impact on the parton distributions. In Fig.~3 we show the gluon
distributions for each of the analyses at several values of $Q^2$. At the starting
scale of $Q^2 = 1\;{\rm GeV}^2$, both NLO and NNLO gluons are negative for very low $x$
with the NLO lying below the NNLO. As $Q^2$ increases, the NNLO gluon evolves
relatively more slowly so that by $Q^2 = 5\; {\rm GeV}^2$ the situation is reversed,
both gluons being positive with the NLO gluon now above the NNLO one. Also shown are
the NNLO gluons from the fits using the extreme allowed forms of the splitting
functions. We note that the spread between these gluons is somewhat reduced,
particularly above $x=10^{-3}$, compared to the range obtained in Ref.~\cite{NNLO} as a
result of the improved reliability of the NNLO splitting functions of Ref.~\cite{VV3}.
We also note that the NNLO gluon is generally closer to the NLO gluon at very low $x$
compared to the case in \cite{NNLO}. This is due to the decrease of the NNLO
quark-gluon splitting function at low $x$ compared to the previous expectation,
therefore allowing more small $x$ gluon at NNLO. As in Ref.~\cite{NNLO} the negative
feature of the gluon at low $x$ and low $Q^2$ is not, in itself, a matter for concern
since it is not a physical observable, being a scheme dependent quantity.

In Fig.~4 we compare the parton distributions found in the NNLO fit to those obtained
in the NLO analysis \cite{MRST2001}.  We plot the NNLO/NLO ratios for the gluon, and
the up and down quark distributions, at $Q^2 = 10$ and $10^4\;{\rm GeV}^2$.  In going
from the NLO to the NNLO analysis, several features of Fig.~4 are worth noting. The
decrease of the quark distribution at high $x$ and the increase at low $x$ (at the
lower $Q^2$ value) reflect the behaviour of the NNLO quark and gluon coefficient
functions respectively. We have already seen above that for $Q^2 > 5\;{\rm GeV}^2$ the
NLO gluon has overtaken the NNLO gluon and this situation remains as the evolution
continues. The rise at large $x$ for the NNLO/NLO gluon ratio is a consequence of the
need for more gluon at high $x$ to fit the Tevatron jet data and also of momentum
conservation -- the positive NNLO quark-gluon splitting function at low $x$ means less
gluon is required here and more is available elsewhere. This enhancement of the gluon
for $x$ above about $0.1$ leads to a particularly improved fit to the NMC DIS data at
NNLO. With the evolution to high $Q^2$, the NNLO effects play less of a role, and the
NNLO coefficient functions have very little impact. However, the relatively smaller
low-$x$ NNLO gluon introduced at low $Q^2$ produces a slower evolution of all partons
at low $x$, which can be seen at $Q^2 = 10^4\;{\rm GeV}^2$.

Computing NNLO predictions for the longitudinal structure function $F_L$ requires a
knowledge of the ${\cal O}(\alpha_S^3)$ coefficient functions. In Ref.~\cite{NNLO} we
used the partial information of the first four even moments together with the $x
\rightarrow 0$ behaviour to estimate analogous expressions to the ${\cal
O}(\alpha_S^3)$ splitting functions of Ref.~\cite{VV3}. Here we continue with that
approach to compare the NNLO prediction for $F_L$ with the NLO one. In fact the two
additional moments lead to the most likely coefficient functions being essentially
unchanged from those used in \cite{NNLO} (a lucky accident), but we note that the band
of uncertainty on these coefficient functions is much larger than on the splitting
functions. Truly predictive results await a more complete analysis. In our recent NLO
analysis \cite{MRST2001} we drew attention to the unphysical behaviour of the NLO
prediction for $F_L$ at low $Q^2$ and very low $x$, shown in Fig.~5, suggesting that
the negative feature may be a hint that NLO cannot provide the entire description in
that region. Now we see that the NNLO prediction is indeed more satisfactory, with the
physical observable $F_L$ remaining positive at $Q^2  = 2\; {\rm GeV}^2$ down to $x =
10^{-5}$. Even this qualitative conclusion is not definite with the present level of
uncertainty. More reliable is the fact that the NNLO prediction is below that at NLO at
high $Q^2$, since this relies less on the coefficient function and more on the gluon
evolution.

The improvement in the description of $F_2$ going from NLO to NNLO is not as dramatic
as that going from LO to NLO. This could be interpreted as evidence for some
convergence as we go to higher orders. One phenomenological question that has been
raised is whether the data which lie below our cut of $W^2 = 12.5 \; {\rm GeV}^2$,
which is imposed to remove possible higher twist contamination, could be consistently
described by NNLO effects. At large $x$ the data below this cut show a marked rise
above the NLO description -- as much as $40 - 50\%$ at low $Q^2$ -- suggesting a
sizeable higher twist contribution. Fig. 6 shows that the differences between NNLO and
NLO at large $x$ are indeed significant, and are always in the correct direction.
However, they are not at the level needed to account for the magnitude of the observed
discrepancies.

We have also investigated the predictions for the $W$ and $Z$ production cross sections
at the Tevatron and the LHC, and compared them with those obtained previously in
Ref.~\cite{NNLO}. For purposes of comparison we have kept all electroweak parameters,
branching ratios {\it etc}. at the same values as in \cite{MRSTWZ,NNLO}. 
The results are shown
in Table~3. As one can see the difference between the 2001 and 2000 predictions is
always of the order of $0.5\%$ or less, showing that neither the new data in the fit
nor the changes in the NNLO splitting functions have led to any significant
changes.\footnote{The predictions in \cite{MRST2001} were made using the `hybrid'
method of  combining NLO partons with the known NNLO coefficient functions.}
However, the error band on the prediction due to the uncertainty on the splitting functions has
been significantly reduced due to the better estimate of the NNLO splitting functions.
At Tevatron energies the difference between the `slow' and `fast' NNLO evolution
predictions is at the per mille level, while at the LHC the difference is still
less than 1\%.

\begin{table}
\caption{Predictions in $nb$ for $W$ and $Z$ production cross sections at the Tevatron
and LHC, compared with those of MRST2000 \cite{NNLO}.}
\begin{center}
\begin{tabular}{|cccc|} \hline
&   & MRST2000 & MRST2001 \\ \hline
{\bf NLO}&   &     &          \\ \hline
Tevatron & $B_{\ell \nu} \cdot \sigma_W$ & 2.39 & 2.41 \\
& $B_{\ell^+ \ell^-} \cdot \sigma_Z$ & 0.219 & 0.220 \\ \hline
LHC & $B_{\ell \nu} \cdot \sigma_W$ & 20.5 & 20.6 \\
& $B_{\ell^+ \ell^-} \cdot \sigma_Z$ & 1.88 & 1.89 \\
\hline {\bf NNLO}&   &    & \\ \hline
Tevatron & $B_{\ell \nu} \cdot \sigma_W$ & 2.51 & 2.50 \\
& $B_{\ell^+ \ell^-} \cdot \sigma_Z$ & 0.230 & 0.230 \\ \hline
LHC & $B_{\ell \nu} \cdot \sigma_W$ & 19.9 & 20.0 \\
& $B_{\ell^+ \ell^-} \cdot \sigma_Z$ & 1.84 & 1.85 \\ \hline
\end{tabular}
\end{center}
\end{table}

In summary, we have have carried out NNLO  and LO global analyses of data which include
the most recent data on DIS from HERA together with constraints from hard scattering
data. Comparisons were made with our recent NLO analysis \cite{MRST2001} for the the
structure functions and for the partons themselves. There is improvement in the general
quality of the description of the data as one goes to higher orders, though the
improvement in the global fit at NNLO is not as great as one might expect, and there is
also the surprising result that the LO partons give an excellent fit to the Tevatron
high-$E_T$ jet data. The partons at NNLO change by up to $15\%$ for $Q^2 \geq 10\; {\rm
GeV}^2$ compared to those at NLO, with the largest change being in the gluon, which
decreases for $x$ less than about $0.05$ but increases for $x>0.1$. The quarks increase
for $x<0.01$ at low $Q^2$ due to the negative gluon coefficient function, but decrease
at lower $x$ for higher $Q^2$ because of the slower evolution due to the smaller gluon.
They also decrease at very high $x$ due to the positive effect of the NNLO coefficient
function.

We note that these are the first generally available sets of partons at NNLO and should
add impetus to the large activity presently under way to compute NNLO corrections to a
variety of hard processes. In particular, the computation of the `missing' NNLO
coefficient functions for the differential Drell-Yan and large-$E_T$ jet cross sections
used in the global fit to constrain the large-$x$ sea and gluon distributions respectively
is now a matter of some urgency.
The NNLO partons discussed ini this paper can be found at {\tt
http://durpdg.dur.ac.uk/hepdata/mrs} together with the recent NLO partons ($\msb$
scheme) from a similar analysis \cite{MRST2001}. For completeness we have also added
the NLO partons for MRST2001 in the DIS scheme and the LO partons discussed in this
paper.

\section{Acknowledgments}

We would like to thank Andreas Vogt, for useful discussions regarding 
NNLO splitting functions. 
RST would like to thank the Royal Society for the award of
a University Research Fellowship. This work was supported in part
by the EU Fourth Framework Programme ``Training and Mobility of
Researchers'', Network ``Quantum Chromodynamics and the Deep
Structure of Elementary Particles'', contract FMRX-CT98-0194(DG 12
- MIHT).

\section*{Appendix}

We use a very similar charm prescription to that adopted in \cite{NNLO}, which is a
generalization of the Thorne-Roberts variable-flavour-number scheme (VFNS) \cite{VFNS}
to NNLO. As explained in \cite{NNLO}, this has to be a {\it model}, since the NNLO
fixed-flavour-number scheme (FFNS) coefficients are not known. In principle we should
be more sophisticated and (i) use a correct matching procedure for the NNLO VFNS
coefficient function (possible at present), (ii) take into account the small
discontinuity of the charm parton distribution at NNLO (also possible at present), and
(iii) also take in account the effect of the NNLO FFNS coefficient function (not
possible at present). Since the full NNLO prescription is missing, and since the NNLO
splitting functions are approximate themselves at present, we adopt a simple procedure
at NNLO. In \cite{NNLO} we just used the massless ${\cal O}(\alpha_S^2)$ gluon
coefficient function multiplied by the velocity of the heavy quark
$\beta=(1-4m_H^2z/(Q^2(1-z))^{0.5}$, which has the correct threshold behaviour but $\to
1$ at high $Q^2$, as the NNLO coefficient function at all $Q^2$. Here $m_H$ is the mass
of the heavy quark. This has the
shortcoming that it leads to a (small) negative contribution at low $Q^2$ and small $x$
since the the massless ${\cal O}(\alpha_S^2)$ gluon coefficient function is negative at
small $x$, whereas we would expect the NNLO FFNS contribution to actually be positive
in this region. Indeed the low $x$, low $Q^2$ charm data are underestimated using our
procedure. Hence in this paper we adopt the minimal simple modification and weight by
$\beta^2$ rather than $\beta$, so that this negative contribution is minimized. This
improves both the fit to charm data and the global fit. This simple approach seems to
us to be sufficient at the present level of theoretical accuracy. Once the splitting
functions are known exactly at NNLO we should improve our NNLO charm model. In
principle one can even produce an approximate NNLO FFNS coefficient function from the
known small $x$ limit \cite{cch} and known threshold logarithms \cite{eric}.

\newpage

\begin{figure}[!h]
\begin{center}
\epsfig{figure=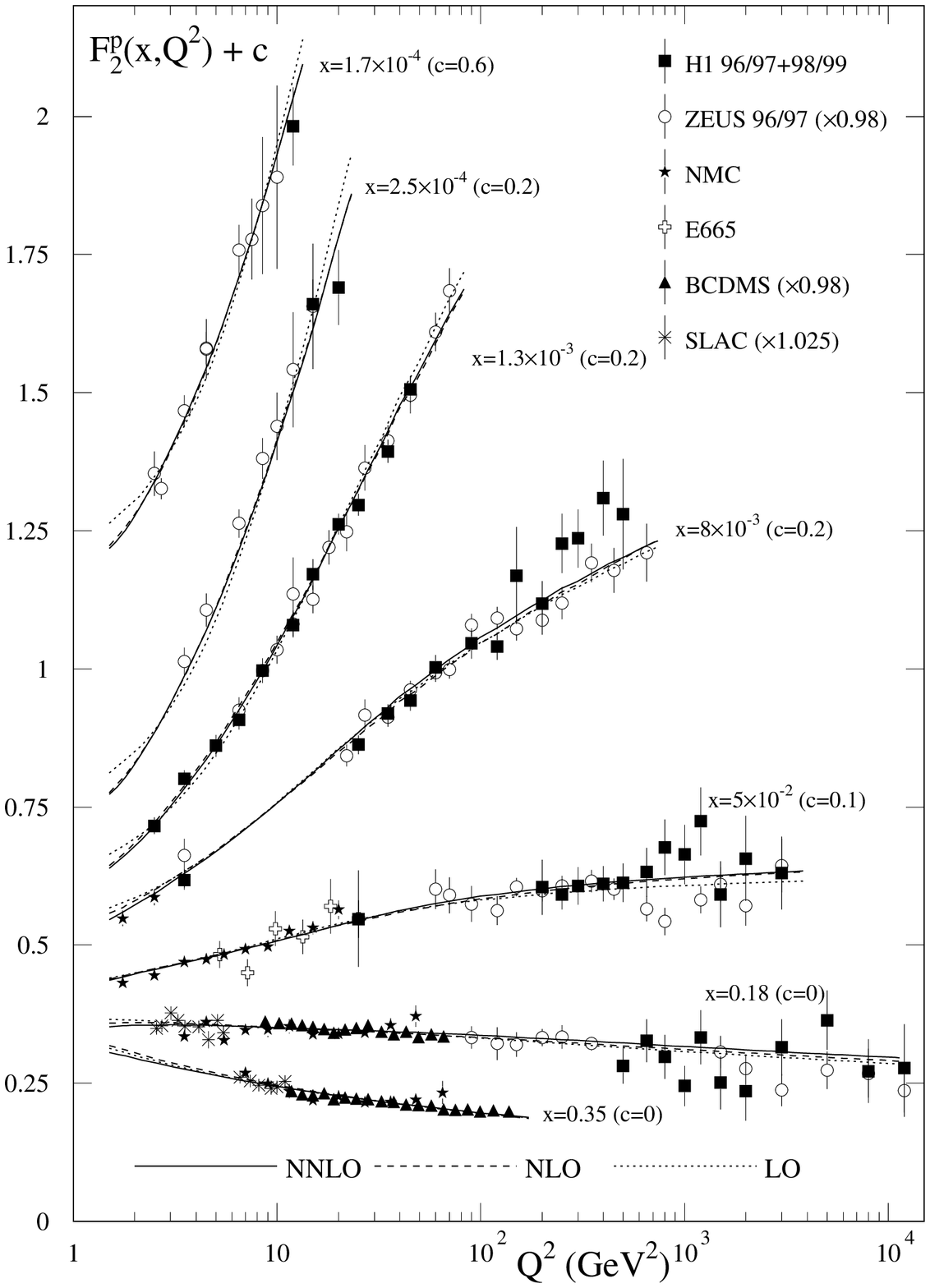,height=8in}
 \caption{The description of data for the $F_2$ structure function
 at a few representative values of $x$ obtained in the LO, NLO
 \cite{MRST2001} and NNLO global analyses.}
\label{Fig1}
\end{center}
\end{figure}

\newpage

\begin{figure}[!h]
\begin{center}
\epsfig{figure=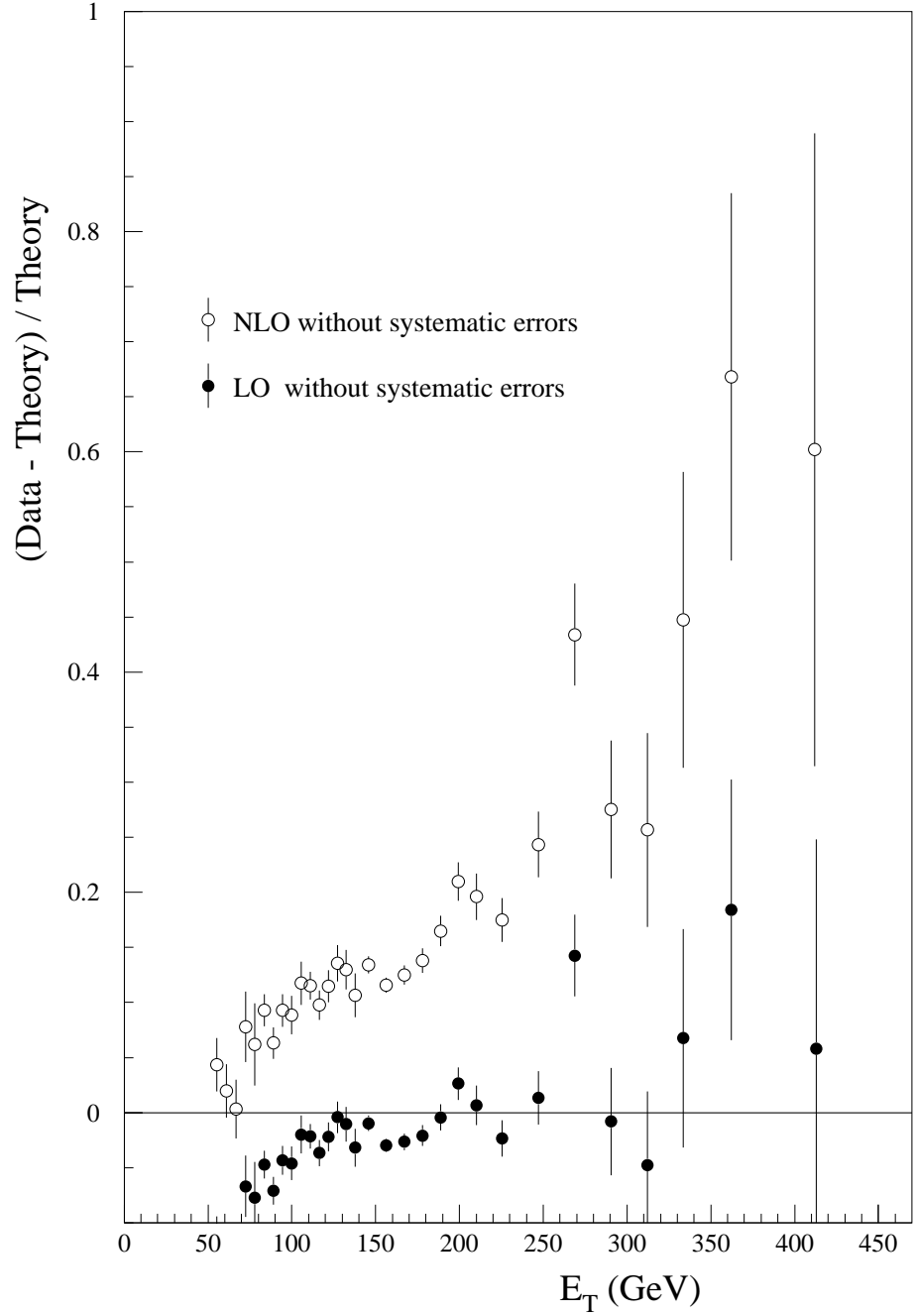,height=8in}
 \caption{The comparison of the quality of the LO MRST and the NLO MRST
fit to the CDF1B high-$E_T$ jet data \cite{CDF}. The open points are the NLO values
before the correlated systematic errors have been considered while the solid points are
those for the LO fit.} \label{Fig2}
\end{center}
\end{figure}

\newpage

\begin{figure}[!h]
\begin{center}
\epsfig{figure=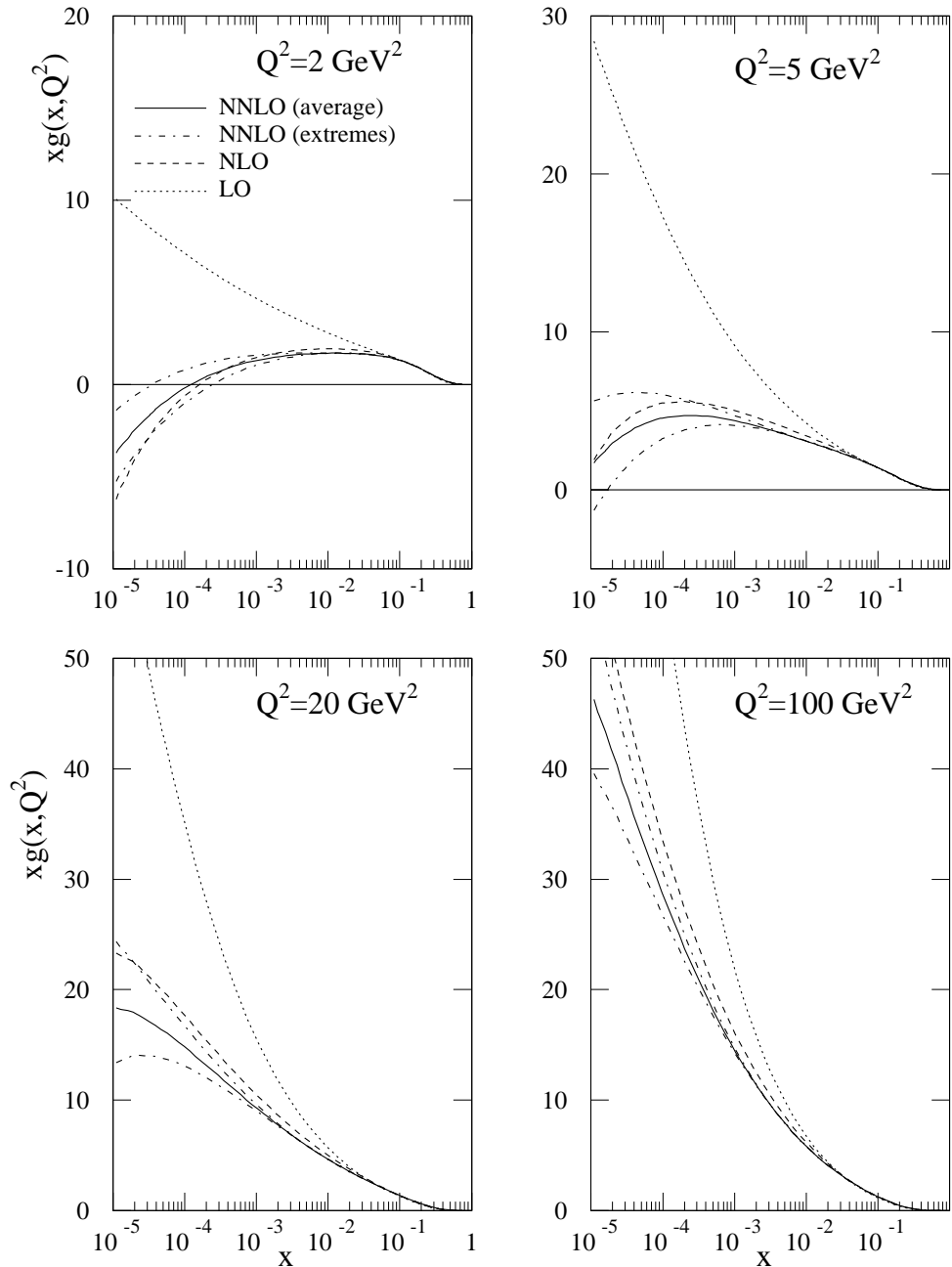,height=8in}
 \caption{Gluon distributions obtained from the LO, NLO and NNLO analyses
 at various values of $Q^2$. The three NNLO gluons result from the
 slow and fast extremes of the splitting functions together with the
 average of these.}
 \label{Fig3}
\end{center}
\end{figure}

\begin{figure}[!h]
\begin{center}
\epsfig{figure=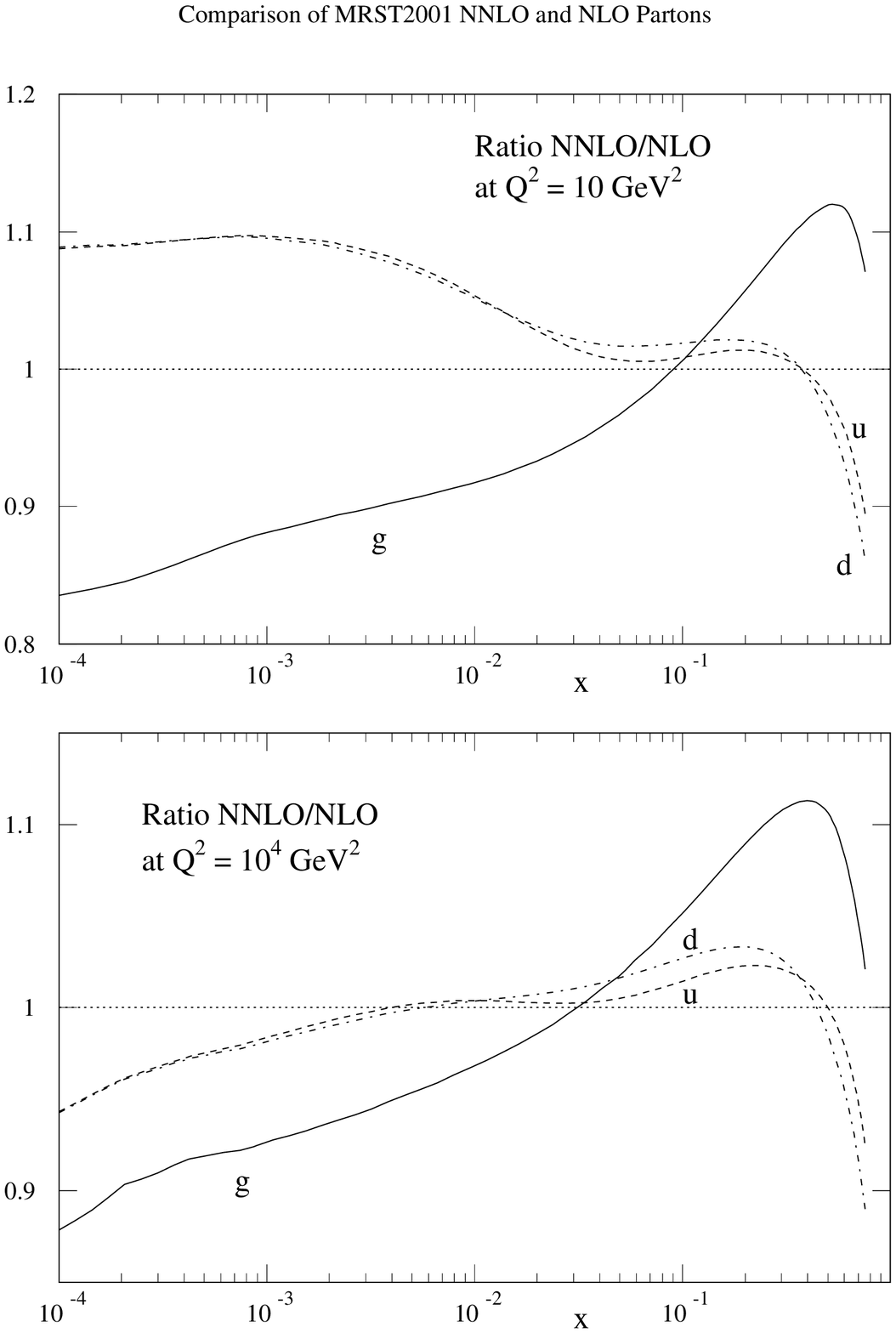,height=8in}
 \caption{A comparison of the gluon, up and down distributions obtained
 in the NNLO analysis with those obtained in the NLO fit \cite{MRST2001}.
 The comparison is shown for two values of $Q^2$.}
 \label{Fig4}
\end{center}
\end{figure}

\begin{figure}[!h]
\begin{center}
\epsfig{figure=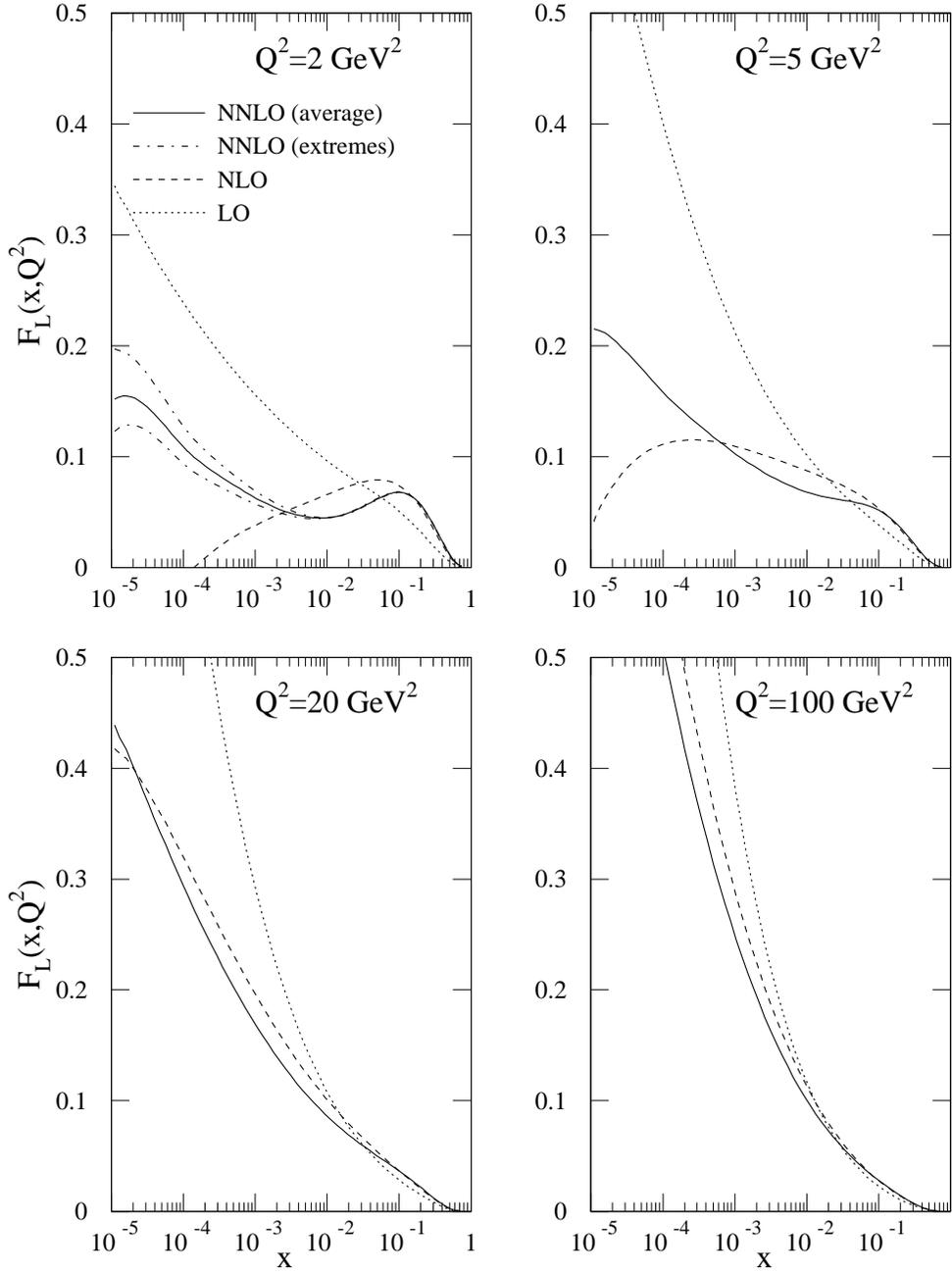,height=8in}
 \caption{Predictions for $F_L$ from the LO, NLO and NNLO partons.}
 \label{Fig5}
\end{center}
\end{figure}

\begin{figure}[!h]
\begin{center}
\epsfig{figure=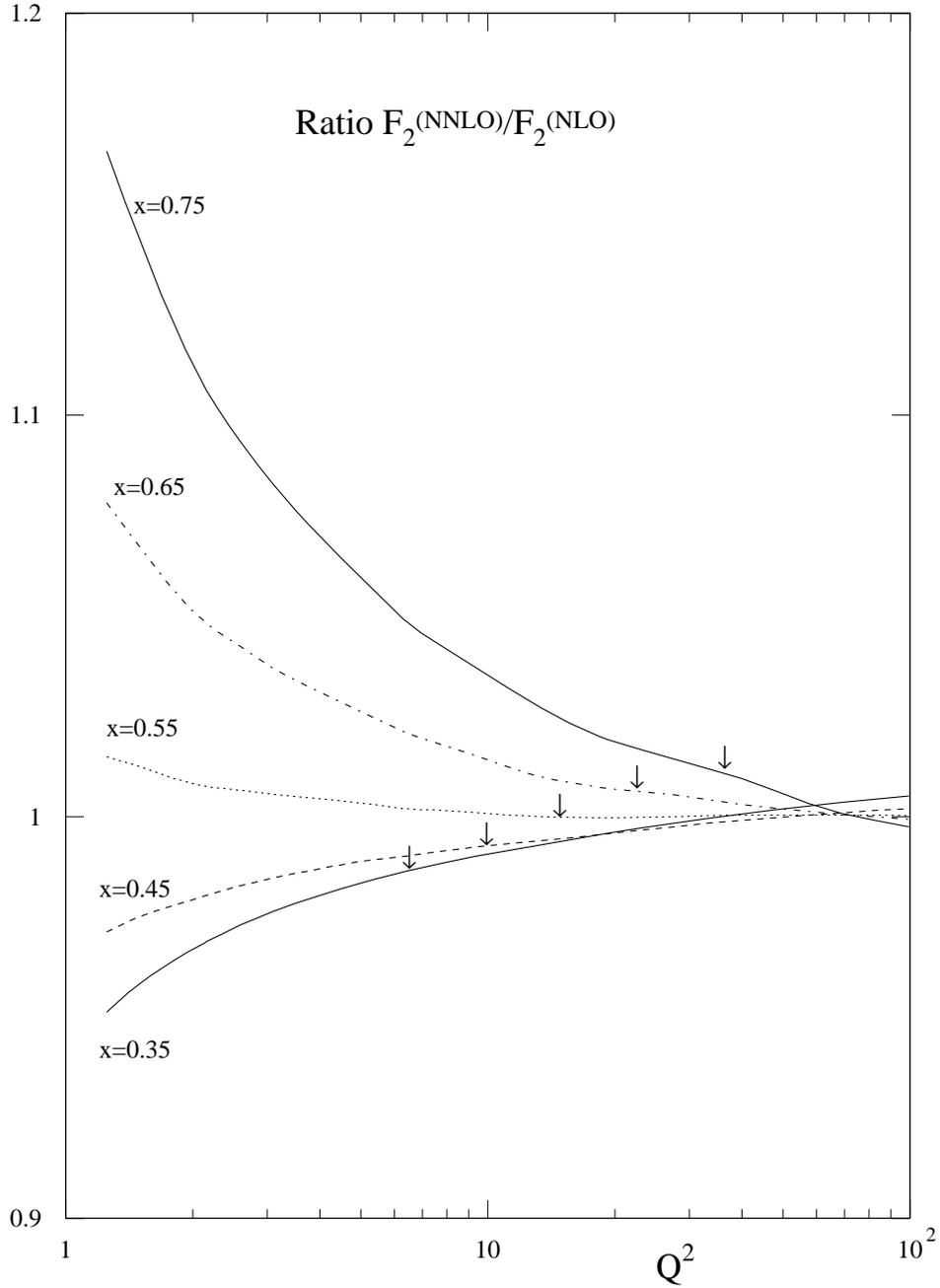,height=8in}
 \caption{Ratio of NNLO and NLO structure function $F_2$ at small and
 large $x$. For large $x$ the position of $W^2 = 12.5$ GeV$^2$ is indicated.}
 \label{Fig6}
\end{center}
\end{figure}

\end{document}